\documentclass{article}

\usepackage[english]{babel}

\usepackage[letterpaper,top=2cm,bottom=2cm,left=3cm,right=3cm,marginparwidth=1.75cm]{geometry}

\usepackage{amsmath}
\usepackage{graphicx}
\usepackage[colorlinks=true, allcolors=blue]{hyperref}
\usepackage{graphicx,times}             
\usepackage[numbers]{natbib}
\usepackage{amssymb,amsmath}
\usepackage{booktabs}
\usepackage{url}
\usepackage{array}
\usepackage{multirow}
\usepackage{supertabular}
\usepackage {rotfloat}
\usepackage{abstract}
\usepackage{color}

\usepackage{authblk}     
\date{}

\begin{document}
\title{Cross calibration of gamma-ray detectors (GRD) of GECAM-C}

\author[1,2]{Yan-Qiu Zhang}
\author[1]{Shao-Lin Xiong \thanks{Corresponding author: xiongsl@ihep.ac.cn}}
\author[1]{Rui Qiao \thanks{Corresponding author: qiaorui@ihep.ac.cn}}
\author[1]{Dong-Ya Guo}
\author[1]{Wen-Xi Peng}
\author[1]{Xin-Qiao Li}
\author[1,2]{Wang-Chen Xue}
\author[1,2]{Chao Zheng}
\author[1,2]{Jia-Cong Liu}
\author[1,2]{Wen-Jun Tan}
\author[1,2]{Chen-Wei Wang}
\author[1,11]{Peng Zhang}
\author[1]{Ping Wang}
\author[1,2,10]{Ce Cai}
\author[1,2,9]{Shuo Xiao}
\author[1]{Yue Huang}
\author[1,2]{Pei-Yi Feng}
\author[1]{Xiao-Bo Li}
\author[1]{Li-Ming Song}
\author[1,4]{Qi-Bin Yi}
\author[1,8]{Yi Zhao }
\author[1,5]{Zhi-Wei Guo}
\author[1]{Jian-Jian He}
\author[1]{Chao-Yang Li}
\author[1]{Ya-Qing Liu}
\author[1]{Ke Gong}
\author[1,6]{Yan-Qi Du}
\author[1]{Xiao-Jing Liu}
\author[1,3]{Sheng-Lun Xie}
\author[1,4]{Guo-Ying Zhao}
\author[1]{Xiao-Yun Zhao}
\author[1,7]{Xiao-Lu Zhang}
\author[1]{Zhen Zhang}
\author[1]{Shi-Jie Zheng}
\author[1]{Jin Wang}
\author[1]{Xiang-Yang Wen}
\author[1]{Zheng-Hua An}
\author[1]{Da-Li Zhang}
\author[1]{Min Gao}
\author[1]{Xi-Lei Sun}
\author[1]{Xiao-Hua Liang}
\author[1]{Sheng Yang}
\author[1]{Jin-Zhou Wang}
\author[1]{Gang Chen}
\author[1]{Fan Zhang}
\renewcommand*{\Authfont}{\normalsize}
\renewcommand\Authands{ and } 
\affil[1]{Key Laboratory of Particle Astrophysics, Institute of High Energy Physics,
Chinese Academy of Sciences, Beijing 100049, China}
\affil[2]{University of Chinese Academy of Sciences, Beijing 100049, China}
\affil[3]{Institute of Astrophysics, Central China Normal University, Wuhan 430079, China}
\affil[4]{School of Physics and Optoelectronics, Xiangtan University, Xiangtan 411105, Hunan, China}
\affil[5]{College of physics Sciences Technology, Hebei University, No. 180 Wusi Dong Road, Lian Chi District, Baoding City, Hebei Province 071002, China}
\affil[6]{Information Science and Technology, Southwest Jiaotong University, Chengdu 610031, China}
\affil[7]{Qufu Normal University, Qufu 273165, China}
\affil[8]{Department of Astronomy, Beijing Normal University, Beijing 100875, China}
\affil[9]{Guizhou Provincial Key Laboratory of Radio Astronomy and Data Processing, Guizhou Normal University, Guiyang 550001, People’s Republic of China}
\affil[10]{College of Physics and Hebei Key Laboratory of Photophysics Research and Application, Hebei Normal University, Shijiazhuang, Hebei 050024, China }
\affil[11]{College of Electronic and Information Engineering, Tongji University, Shanghai 201804, China }
\renewcommand*{\Affilfont}{\small}

\maketitle

\vspace{-4.0em}
\renewcommand{\abstractname}{\LARGE Abstract\\}
\begin{abstract}

{\normalsize The gamma-ray detectors (GRDs) of GECAM-C onborad SATech-01 satellite is designed to monitor gamma-ray transients all over the sky from 6 keV to 6 MeV. The energy response matrix is the key to do spectral measurements of bursts, which is usually generated from GEANT4 simulation and partially verified by the ground calibration. In this work, energy response matrix of GECAM-C GRD is cross-calibrated with \emph{Fermi}/GBM and \emph{Swift}/BAT using a sample of Gamma-Ray Bursts (GRBs) and Soft Gamma-Ray Repeaters (SGRs). The calibration results show there is a good agreement between GECAM-C and other reasonably well calibrated instrument (i.e. \emph{Fermi}/GBM and \emph{Swift}/BAT). We also find that different GRD detectors of GECAM-C also show consistency with each other. All these results indicate that GECAM-C GRD can provide reliable spectral measurements.\\
\renewcommand{\baselinestretch}{1.5}
}
\end{abstract}

\section{INTRODUCTION}
\begin{figure}
\vspace{-2.0em}
   \centering
   \includegraphics[width=\textwidth, angle=0]{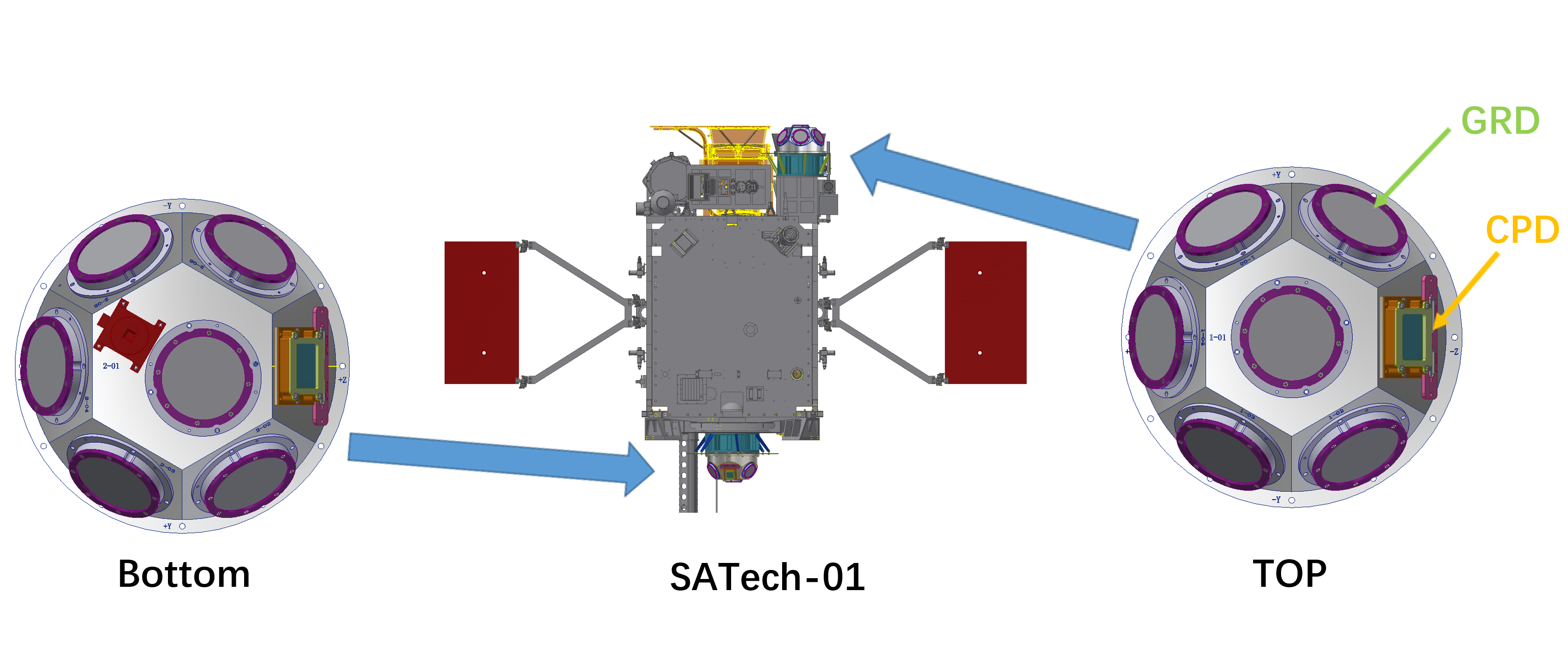}
   \caption{The installation position of GECAM-C and the installation location of detectors on GECAM-C.}
   \label{fig:GECAM-C_installation}
\end{figure}

GECAM-C \cite{zhang2023SiPM} (also called High Energy Burst Researcher, HEBS) is one of the payloads on the New Space Technology Test Satellite (SATech-01) launched on July 27,2022 \cite{qie2021study}. It consists of two hemispherical domes, which follows the design of the Gravitational wave high-energy Electromagnetic Counterpart All-sky Monitor (GECAM) \cite{li2021technology}. These two domes are installed on the top and bottom of SATech-01 satellite. There are six gamma-ray detectors (GRDs) \cite{an2022design} and one charge particle detector (CPD) \cite{xu2022design} on each dome, therefore, GECAM-C has a total of 12 GRDs and two CPDs, as shown in Fig \ref{fig:GECAM-C_installation}.

For each dome, there are two types of GRDs: the detectors composed of LaBr3 crtystals (LGRDs) and the detectors composed of NaI crystals (NGRDs). Both types of GRD are readout by SiPM array \cite{lv2018low}\cite{zhang2022dedicated}\cite{zhang2023SiPM}. There are two types of readout case: the single channel and the dual channels (i.e. High-Gain and Low-Gain). All the basic information are listed in the Table \ref{The basic information of GECAM-C}. The designed energy range for LGRD is 10-332 keV for High-Gain channel and 122-4500 keV for Low-Gain channel, and for NGRD is 10-355 keV for High-Gain channel and 122-5889 keV for Low-Gain channel. But for the NGRD with single gain channel, its designed detection energy range is 10 to 700 keV.

The energy response matrix of the GECAM-C GRDs are generated from GEANT4 simulation, similar as the case for GECAM-A and GECAM-B \cite{guo2020energy}. The simulation results are partially validated using the ground calibration tests \cite{zheng2023GroundCal}. These help us to establish the response matrix for GECAM-C. In this paper, we focus on the cross-calibration of the response matrix through joint analysis with \emph{Fermi}/GBM and \emph{Swift}/BAT.

\begin{center} 
\begin{sidewaystable}
\caption{The basic information of GECAM-C}
\begin{tabular*}{\hsize}{@{}@{\extracolsep{\fill}}cccccc@{}}
\hline
\midrule


  & \multirow{2}*{Detectors type} & \multirow{2}*{Electronic readout channel} & \multirow{2}*{Detector number} &   \multicolumn{2}{c}{Detector orientations} \\
  \cline{5-6}
  & & & & $\theta$ ($^{\circ}$) & $\phi$ ($^{\circ}$)\\
  
\midrule

 \multirow{7}*{Top dome cabin}  
& LGRD & High-Gain and Low-Gain & GRD01 & 0 & 0 \\
& NGRD & High-Gain and Low-Gain & GRD02 & 60 & 210 \\
& LGRD & High-Gain and Low-Gain & GRD03 & 60 & 150 \\
& NGRD & High-Gain and Low-Gain & GRD04 & 60 & 90 \\
& LGRD & High-Gain and Low-Gain & GRD05 & 60 & 30 \\
& NGRD & Single & GRD06 & 60 & 330 \\
& CPD & Single & 1-01C & 60 & 270 \\
\midrule
 \multirow{7}*{Bottom dome cabin}  
& LGRD & High-Gain and Low-Gain & GRD07 & 180 & 0 \\
& NGRD & High-Gain and Low-Gain & GRD08 & 120 & 150 \\
& LGRD & High-Gain and Low-Gain & GRD09 & 120 & 210 \\
& NGRD & High-Gain and Low-Gain & GRD10 & 120 & 270 \\
& LGRD & High-Gain and Low-Gain & GRD11 & 120 & 330 \\
& NGRD & Single & GRD12 & 120 & 30 \\
& CPD & Single & 2-01C & 120 & 90 \\

\bottomrule\label{The basic information of GECAM-C}
\end{tabular*}
\footnotesize{The Detectors type means GRD composed of two crystals: LaBr3 and NaI; \\High-Gain and Low-Gain mean the two read out channels; \\Detector orientations mean the normal pointing of GRDs in the GECAM-C payload coordinate system.
}\\
\end{sidewaystable}
\end{center}

\section{JOINT SPECTRUM FITTING}
Cross calibration is implemented by joint spectrum fitting of burst events that are observed by GECAM-C and other well-calibrated instruments (here we choose Fermi/GBM and Swift/BAT to do cross calibration). The correctness of the GECAM-C response matrix can be judged from the consistency of the fitting results. This method has been used in cross calibration of many instruments\cite{luo2020calibration}\cite{sakamoto2011spectral} . 

In this paper, we will also multiply GECAM-C by a constant factor and the constant factor of reference instruments (i.e. Fermi/GBM or Swift/BAT) is fixed to one, and judge whether the response matrix of GECAM-C can well describe the scientific data of GECAM-C.
In this section we will introduce the instruments used to perform the cross calibration as well as the analytical process and results.

\subsection{Instruments}

\subsubsection{\emph{Fermi}/GBM}
As one of the two instruments on the \emph{Fermi Gamma-ray Space Telescope}, Fermi Gamma-ray Burst Monitor (\emph{Fermi}/GBM) is composed of NaI(Tl) detectors (8-1000 keV) and BGO detectors (~200 keV to ~40 MeV) \cite{meegan2009fermi}. From the detection energy range of \emph{Fermi}/GBM, it is very suitable for the cross-calibration work with GECAM-C, because they have the same detection energy band. And the \emph{Fermi}/GBM has been reasonably well calibrated \cite{stamatikos2009cross}\cite{tierney2011spectral}\cite{von2009using}. 

\subsubsection{\emph{Swift}/BAT}
The Swift Burst Alert Telescope (\emph{Swift}/BAT) is one of the instruments on the Swift Gamma-ray Burst Explorer \cite{gehrels2004swift}. The detector of \emph{Swift}/BAT is made of CdZnTe (CZT), and the detectable energy range is 15–150 keV \cite{suzuki2003hard}, which can be used to cross-calibrate GECAM-C response matrix at the low-energy band. It is worth mentioning that the Swift burst usually has accurate location, which is very helpful for cross calibration.

\begin{center} 
\begin{table*}
\setlength{\tabcolsep}{0.009pt}
\caption{Information of GRBs used for calibration}
\begin{tabular*}{\hsize}{@{}@{\extracolsep{\fill}}ccc@{}}
\hline
\midrule
GRB Name & GRB 220921A & GRB 220930A\\
\midrule
Trigger Time (UTC) & 2022-09-21T11:05:59.070 & 2022-09-30T11:11:52.000\\
(Ra[$^{\circ}$],Dec[$^{\circ}$],error[$^{\prime \prime}$]) & (66.47, -40.41, 3.8) & (64.72, 13.35, 2.2)\\
Time Interval (s) & [0,15]& [-2,8]\\
($\theta$[$^{\circ}$],$\phi$[$^{\circ}$]) & (29.69, 34.07) & (141.52, 115.48)\\
Joint Instrument & \emph{Fermi}/GBM (n2,na,b0,b1) & \emph{Swift}/BAT\\
GECAM-C Detectors & GRD 01,04,05 & GRD 07,08,09\\
\bottomrule\label{Basic information of the GRB sample}
\end{tabular*}
\footnotesize{The time interval for joint calibration represent the selected time relative to \emph{Fermi}/GBM trigger time;\\
Ra and Dec are the location of the GRB burst;\\
$\theta$[$^{\circ}$],$\phi$[$^{\circ}$] denote the incident angle of the source in the GECAM-C payload coordinate system;\\
GECAM-C Detectors mark the three brightest GRD detectors of GECAM-C used for calibration};
\end{table*}
\end{center}

\begin{figure}[H]
\vspace{-4.0em}
   \centering
   \includegraphics[width=0.9\textwidth, angle=0]{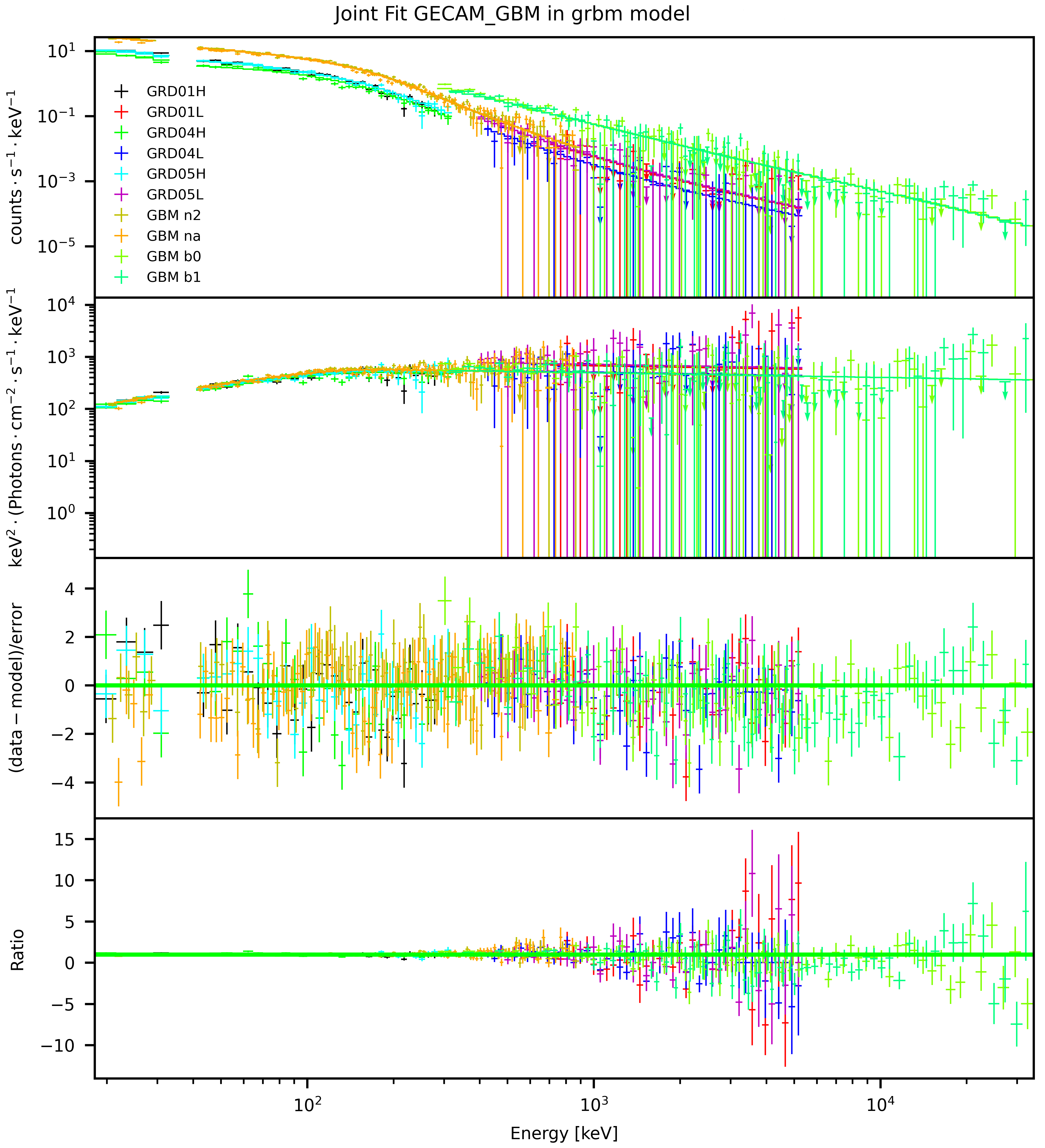}
   \caption{Spectrum fitting results for GRB 220921A; The first subfigure is the net spectrum of the deposition (data points with errors) and the result after the convolution response of the model (step solid line), and it can be seen that the model fits the data; The second subgraph is the energy flux $\rm \nu \mathcal {F}_{\nu}$; The third subgraph is the residuals in terms of sigmas with error bars of size one; The fourth subgraph is the ratio of the data to the model, which means when the ratio is close to 1 (thick green line), the fitting result is very good.}
   \label{fig:Joint_fit_GECAM_GBM_grbm_phase1_False}
\end{figure}

\subsection{Jonit Analysis of GRBs}
\subsubsection{GRB 220921A}
The trigger time of GRB 220921A\footnote{\url{https://gcn.gsfc.nasa.gov/other/220921A.gcn3}} is at 2022-09-21T11:05:59.070 (UTC), and this is the \emph{Fermi}/GBM trigger time. The light curves detected by GECAM-C are shown in Fig \ref{fig:GRB220921A_GECAM-C}. This burst is located at RA=66.47 deg, Dec=-40.41 deg, with an error of 3.8 arc seconds. It is worth mentioning that accurate location is beneficial to cross calibration, which can rule out the effect of inaccurate location on the response matrix.
In addition, we used this burst to cross-calibrate \emph{Fermi}/GBM and GECAM-C.

For this GRB, we choose the two NaI(Tl) of \emph{Fermi}/GBM with highest count rate, respectively are n2 and na. According to the official example of an burst analysis: when we have chosen NaI(Tl) detectors between 0 and 5, we also select BGO0, and if we have chosen NaI detectors 6-11, we would use BGO1\footnote{\url{https://fermi.gsfc.nasa.gov/ssc/data/analysis/scitools/rmfit_tutorial.html}}. Therefore, b0 and b1 all will be selected for GRB 220921A. In addition, as we can see from Fig \ref{fig:GRB220921A_GECAM-C}, GRD01, GRD04 and GRD05 are the three brightest detectors. So we will use these detectors for joint calibration. All of this information can be found in the Table \ref{Basic information of the GRB sample}, including the following GRB 220930A  \footnote{\url{https://gcn.gsfc.nasa.gov/other/220930A.gcn3}}.

\begin{figure}
   \centering
   \includegraphics[width=1.15\textwidth, angle=0]{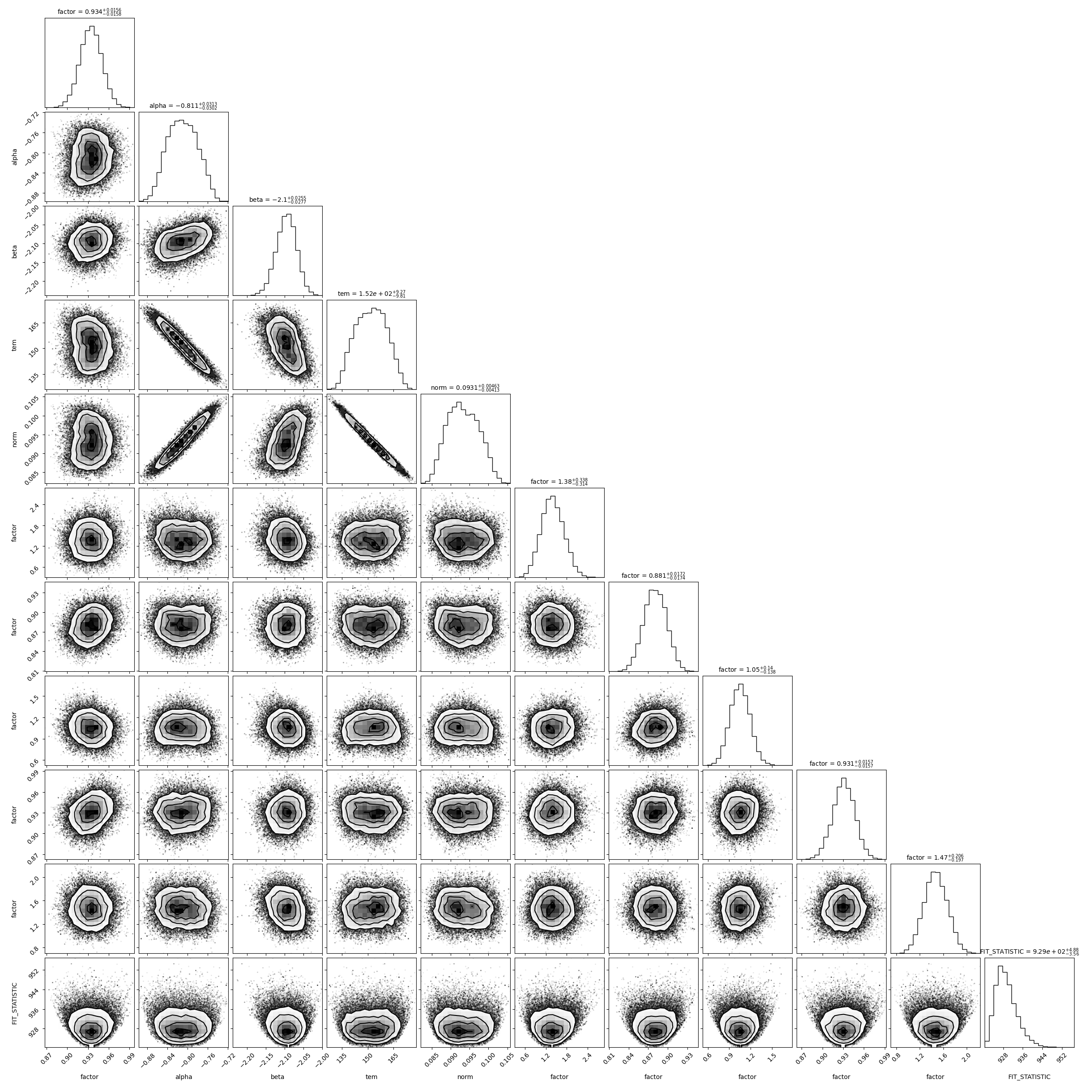}
   \caption{Corner plot of MCMC results for the spectrum parameters; Hgh-Gain and Low-Gain of GECAM-C separate untie; The constant factor of \emph{Fermi}/GBM is fixed to one; As we can see from this figure, all the spectrum parameters are very well constrained.}
   \label{fig:Corner_of_GECAM_GBM}
\end{figure}

In addition, we have tried some spectrum models and the best model for this GRB is Band model\cite{band1993batse}. Different fitting strategies are implemented which include: (1) High-Gain and Low-Gain fit together; (2) High-Gain and Low-Gain separate untie. The result using the second fitting method is shown in the Fig \ref{fig:Joint_fit_GECAM_GBM_grbm_phase1_False}, the residual plot fluctuates around zero and the ratio plot fluctuates around one, which all mean the Band model fits the data points very well. The MC corner plot of the model parameters is presented in Fig \ref{fig:Corner_of_GECAM_GBM}, we can see that the parameters converge very well. In conclusion, it is certain that the results of joint fitting are very good. The fitted parameter results in both ways will be listed in the Table \ref{Model fit results for GRB 220921A fit together} and Table \ref{Model fit results for GRB 220921A untie}, respectively, and we can see that all the spectral parameters of the two kinds of fitting strategies are consistent within the error. From the table we also can see that the constant factors describing the consistency of the GECAM-C response matrix are essentially close to one but with some statistical fluctuations. From the above various results we can draw a conclusion, the response matrixs of GECAM-C can basically describe the scientific data well.
 
\subsubsection{GRB 220930A}
GRB 2209030A triggered at T0: 2022-09-30T11:11:52.000 (UTC), \emph{Swift}/BAT also was triggered by it and provided a precise location (ra: 64.72 deg, dec:13.35 deg, with with an error of 2.2 arc second), which means we can use this burst to do a joint analysis with \emph{Swift}/BAT to calibrate the correctness of the low-energy range of the GECAM-C response matrixs. 

GECAM-C light curves of this GRB are shown in Fig \ref{fig:GRB220930A_GECAM-C}. From this figure we can see, this burst is slightly weaker than GRB 220921A due to a larger incident angle. We also calibrated three of the brightest detectors: GRD07, GRD08 and GRD09, with the time interval (T0+[-2,8] s) we choose is shown in Fig \ref{fig:GECAM-C_lightcurve_GRB220930A}, this is the brightest time period of the burst.

\begin{figure}[H]
   \centering
   \includegraphics[width=0.8\textwidth, angle=0]{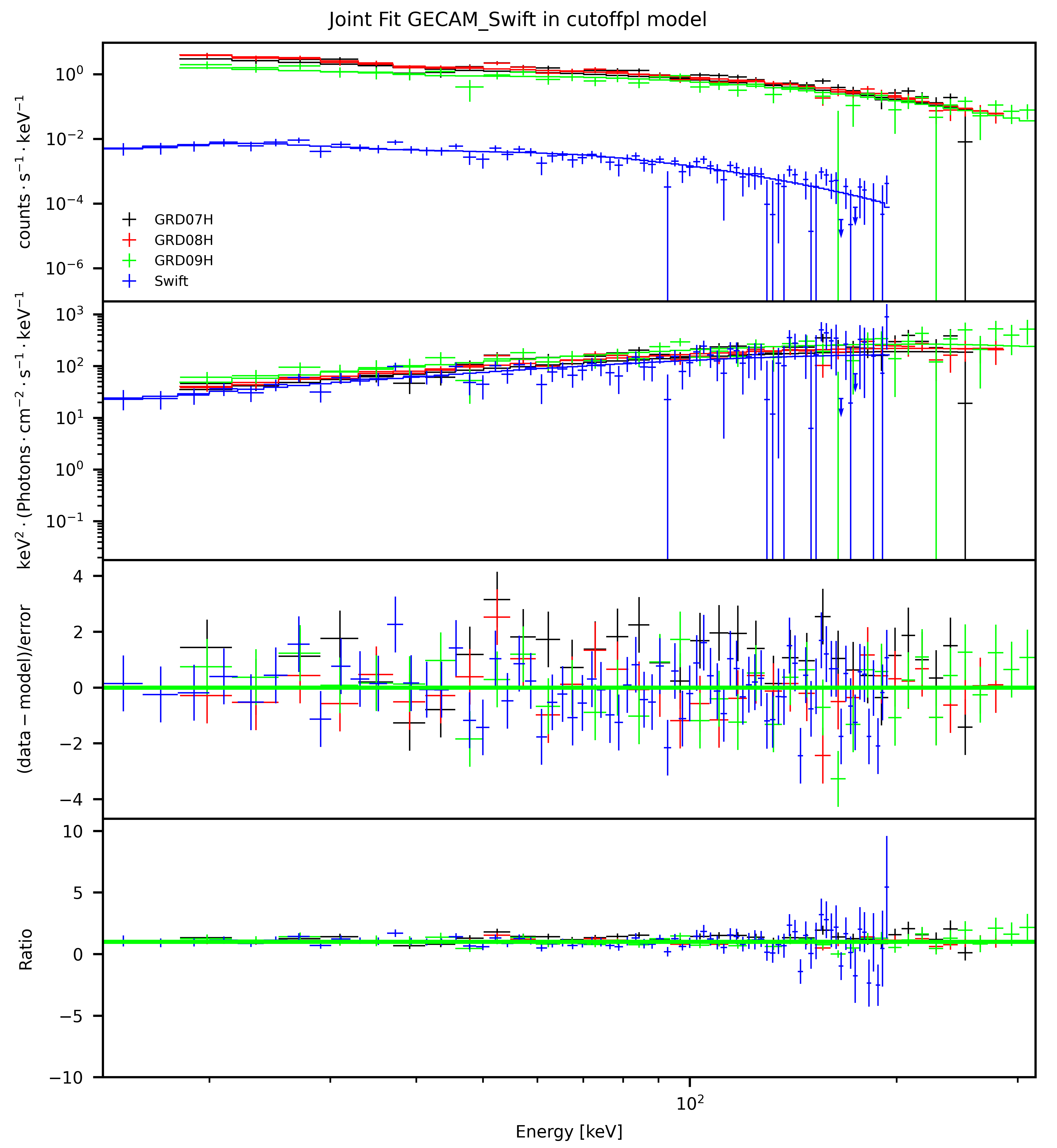}
   \caption{Spectrum fitting results for GRB 220930A; The meaning for each subgraph is the same as the Fig \ref{fig:Joint_fit_GECAM_GBM_grbm_phase1_False}; You can see that for High-Gain the number of photons is sufficient, and for the black (GRD07H), red (GRD08H) and blue (GRD09H) data points the fitting residuals are close to zero, which mean the fitting results are very good.}
   \label{fig:Joint_fit_GECAM_Swift_cutoffpl_phase1_True}
\end{figure}

One thing to notice is that for GRB 220930A, there are very few photons at high energies, and the statistical fluctuations are very large for the Low-Gain of GECAM-C, so the constant factor cannot be limited. In other words, this burst can only be used to calibrate the High-Gain of GECAM-C. As shown in Fig \ref{fig:Joint_fit_GECAM_Swift_cutoffpl_phase1_True}, there is clearly a cut-off in the spectrum, and fortunately using only \emph{Swift}/BAT and GECAM-C High-Gain data can limit the spectral Model, and the fit results can be found in Table \ref{Model fit results for GRB 220930A fit with High-Gain}. From the Fig \ref{fig:Joint_fit_GECAM_Swift_cutoffpl_phase1_True} and Table \ref{Model fit results for GRB 220930A fit with High-Gain}, we can conclude that the consistency between GECAM-C and \emph{Swift}/BAT is pretty good because the constant factor of GECAM-C is 1 within the error range and meanwhile spectrum parameters are well constrained.

From the calibration results of this burst, GECAM-C High-Gain response matrix is a good description of its scientific data and is in good agreement with \emph{Swift}/BAT that has been calibrated.
 
 \begin{center} 
\begin{table*}
\vspace{-4.0em}
\setlength{\tabcolsep}{0.009pt}
\caption{Model fit results for GRB 220921A High-Gain and Low-Gain fit together}
\begin{tabular*}{\hsize}{@{}@{\extracolsep{\fill}}cccccc@{}}
\hline
\midrule
$\alpha$ & $\beta$ &  $E_{\rm cut}$ (keV) &  GRD01 & GRD04 & GRD05\\
\midrule
 $-0.80^{+0.03}_{-0.04}$  &$-2.07^{+0.02}_{-0.03}$  &$148.50^{+11.32}_{-8.87}$  &$0.93^{+0.02}_{-0.01}$  &$0.88^{+0.02}_{-0.01}$  &$0.94^{+0.02}_{-0.01}$  \\
\bottomrule\label{Model fit results for GRB 220921A fit together}
\end{tabular*}
\footnotesize{GRD01,04,05 represent the constant factors of the respective detectors\\}
\end{table*}
\vspace{-2.0em}
\end{center}

\begin{center} 
\begin{table*}
\vspace{-2.0em}
\setlength{\tabcolsep}{0.009pt}
\caption{Model fit results for GRB 220921A with High-Gain and Low-Gain separate untie.}
\begin{tabular*}{\hsize}{@{}@{\extracolsep{\fill}}ccccccccc@{}}
\hline
\midrule
\multirow{2}*{$\alpha$} & \multirow{2}*{$\beta$} &  \multirow{2}*{$E_{\rm cut}$ (keV)} &  \multicolumn{2}{c}{GRD01} & \multicolumn{2}{c}{GRD04} & \multicolumn{2}{c}{GRD05}\\
\cline{4-9}
 & & & High & Low & High & Low & High & Low\\
\midrule
 $-0.81^{+0.03}_{-0.03}$  &$-2.09^{+0.02}_{-0.03}$  &$151.38^{+10.03}_{-9.15}$  &$0.93^{+0.02}_{-0.01}$  &$1.36^{+0.36}_{-0.30}$  &$0.88^{+0.02}_{-0.02}$  &$1.03^{+0.16}_{-0.12}$  &$0.93^{+0.02}_{-0.02}$  &$1.45^{+0.22}_{-0.18}$ \\
\bottomrule\label{Model fit results for GRB 220921A untie}
\end{tabular*}
\footnotesize{GRD01,04,05 represent the constant factors of the respective detectors:  High and Low represent High-Gain and Low-Gain read out channels\\}
\end{table*}
\end{center}

\begin{center} 
\begin{table*}
\vspace{-2.0em}
\setlength{\tabcolsep}{0.009pt}
\caption{Model fit results for GRB 220930A}
\begin{tabular*}{\hsize}{@{}@{\extracolsep{\fill}}cccccc@{}}
\hline
\midrule
$\alpha$ & $E_{\rm cut}$ (keV) & norm &  GRD07 & GRD08 & GRD09\\
\midrule
$-0.90^{+0.10}_{-0.11}$  &$209^{+59}_{-41}$  &$1.62^{+0.80}_{-0.48}$  &$1.20^{+0.21}_{-0.31}$  &$1.06^{+0.18}_{-0.27}$  &$1.27^{+0.23}_{-0.32}$  \\
\bottomrule\label{Model fit results for GRB 220930A fit with High-Gain}
\end{tabular*}
\footnotesize{GRD07,08,09 represent the constant factors of the respective detectors\\}
\end{table*}
\vspace{-2.0em}
\end{center}

\subsection{Jonit Analysis of SGRs}
Bright SGRs can also be used in cross-calibration as GRBs. However, since the energy spectrum of SGR is relatively soft, so it can only be used to calibrate the low energy range. In the case of GECAM-C, it is High-Gain.

\begin{figure}[H]
   \centering
   \includegraphics[width=0.65\textwidth, angle=0]{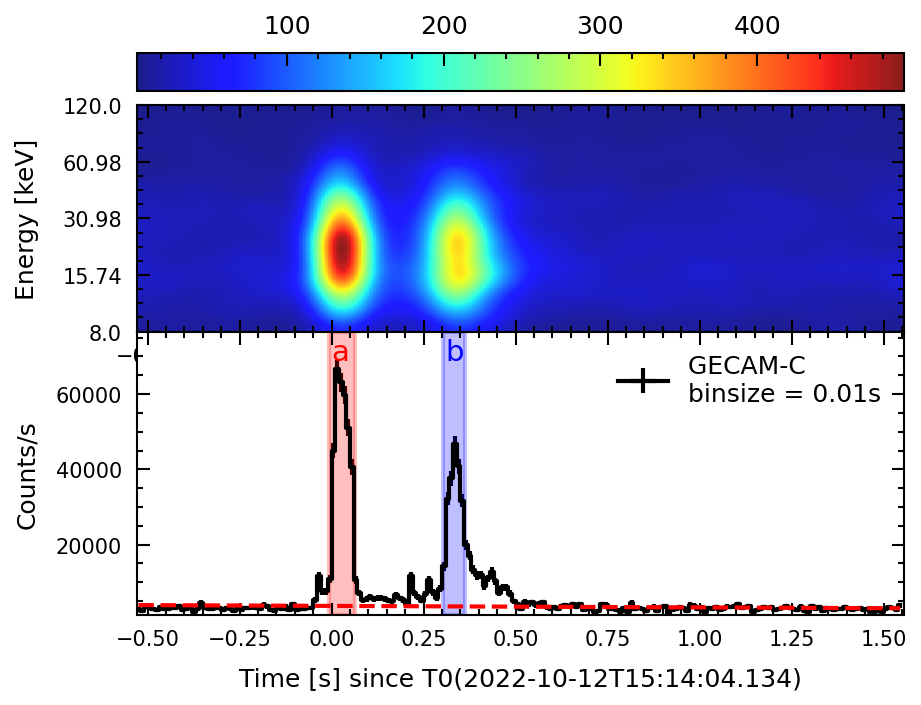}
   \caption{GECAM-C total light curve of SGR22101215 and the time interval selection for joint calibration; The first subgraph is the thermodynamic diagram of the signal detectors (GRD 07,08,09,10,11,12), and the shaded intervals of the second subgraph are the for time ranges (T0+[0.01,0.08],[0.32,0.38] s) of the joint energy spectrum fitting.}
   \label{fig:GECAM-C_lightcurve_SGR22101215}
\end{figure}

\begin{figure}[H]
   \centering
   \includegraphics[width=0.65\textwidth, angle=0]{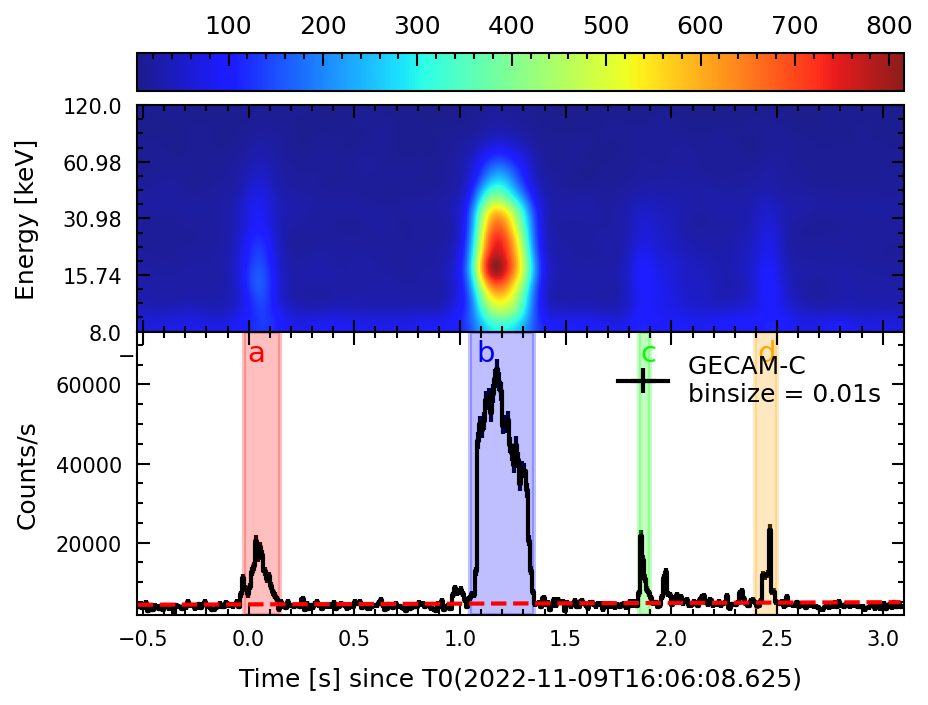}
   \caption{GECAM-C total light curve of SGR22110916 and the time interval selection for joint calibration; The first subgraph is the thermodynamic diagram of the signal detectors (GRD 01,05,06,11,12), and the shaded intervals of the second subgraph are the time ranges (T0+[-0.04,0.13] s,[1.03,1.33] s,[1.83,1.88] s,[2.38,2.48] s) of the joint energy spectrum fitting.}
   \label{fig:GECAM-C_lightcurve_SGR22110916}
\end{figure}

Since the GECAM-C was launched not long ago, there are not enough GRB samples for calibration, so we single out two bright SGRs (SGR 1935+2154) with \emph{Fermi}/GBM joint observation for calibration, and the lightcurves are shown in Fig \ref{fig:GECAM-C_lightcurve_SGR22101215} and Fig \ref{fig:GECAM-C_lightcurve_SGR22110916} respectively. In addition, for each SGR we split several time segments for fitting. At the same time, because SGR is a short-time scale outbreak, therefore, before we do the spectrum fitting we will correct the optical travel difference between the two satellites first. The exact values are shown in the Table \ref{Basic information of the SGRs sample}.

As mentioned before, SGR samples can only be used to calibrate the low-energy part of the instruments, therefore, for this sample, we only selected two relatively bright NaI(Tl) detectors of \emph{Fermi}/GBM and the High-Gain of GECAM-C for joint data analysis. Additional details about the SGRs are listed in the Table \ref{Basic information of the SGRs sample}.

\subsubsection{SGR 22110916}
We selected four time segments of SGR 22110916 for the joint data analysis, and the results are organized in the Table \ref{Model fit results for SGR 22110916 fit with High-Gain}. In addition, the spectrum fitting results are shown in the Fig \ref{fig:GECAM-C_spectrum_result_of_SGR22110916}. Even though the statistical fluctuation of the data points is more obvious because the number of photons is a little less than that of GRB, the overall fitting residuals are around zero, which mean that the fitting results are very good. What's more, from the constant factor in the Table \ref{Model fit results for SGR 22110916 fit with High-Gain}, it's almost always 1, all these indicate that the GECAM-C GRD01,05 detectors are well calibrated for the High-Gain.

\subsubsection{SGR 22101215}
As with SGR 22110916, we also performed a time-resolved spectrum analysis of this burst. The results are presented in Fig \ref{fig:GECAM-C_spectrum_result_of_SGR22101215} and Table \ref{Model fit results for SGR 22101215 fit with High-Gain}. Even though it has similar brightness to SGR 22110916, the duration of the time segment is shorter, so the statistical fluctuation is going to be a little bit larger. However, within the margin of error it can be said that GECAM-C GRD07, 09, 10 detectors and \emph{Fermi}/GBM have relatively good agreement, at the same time, the three detectors also have good agreement with each other.

\begin{center} 
\begin{table*}
\setlength{\tabcolsep}{0.009pt}
\caption{Information of SGRs used for calibration}
\begin{tabular*}{\hsize}{@{}@{\extracolsep{\fill}}ccc@{}}
\hline
\midrule
GRB Name & GRB SGR22101215 & SGR22110916\\
\midrule
Trigger Time (UTC) & 2022-10-12T15:14:04.134 & 2022-11-09T16:06:08.625\\
(Ra[$^{\circ}$],Dec[$^{\circ}$],error[$^{\prime}$]) & (293.743, +21.896, 3) & (293.743, +21.896, 3)\\
Time Interval (s) & [0.01,0.08],[0.32,0.38] & [-0.04,0.13], [1.03,1.33], [1.83,1.88], [2.38,2.48]\\
($\theta$[$^{\circ}$],$\phi$[$^{\circ}$]) & (141.62,231.78) & (63.16,346.62)\\
Joint Instrument & \emph{Fermi}/GBM (n2,na) & \emph{Fermi}/GBM (n0,n1)\\
GECAM-C Detectors & GRD 07,09,10 & GRD 01,05\\
Aberration & -0.0177 & 0.0179\\
\bottomrule\label{Basic information of the SGRs sample}
\end{tabular*}
\footnotesize{The trigger time is the \emph{Fermi}/GBM trigger time;\\
The time interval for joint calibration represent the selected time relative to the trigger time;\\
Ra and Dec are the location of the GRB burst;\\
$\theta$[$^{\circ}$],$\phi$[$^{\circ}$] denote the incident angle of the source in the GECAM-C payload coordinate system;\\
GECAM-C Detectors mark the three brightest GRD detectors of GECAM-C used for calibration;\\
If the aberration is less than zero, it means that GECAM-C triggered later than \emph{Fermi}/GBM};
\end{table*}
\end{center}

\begin{center} 
\begin{table*}
\setlength{\tabcolsep}{0.009pt}
\caption{Model fit results for SGR 22110916}
\begin{tabular*}{\hsize}{@{}@{\extracolsep{\fill}}cccccc@{}}
\hline
\midrule
Time range (s) &$\alpha$ & $E_{\rm cut}$ (keV) & norm &  GRD01 & GRD05\\
\midrule
(-0.04,0.13) & $-0.48^{+0.11}_{-0.09}$  &$17.98^{+0.90}_{-0.88}$  &$26.36^{+6.38}_{-5.98}$  &$1.05^{+0.10}_{-0.03}$  &$1.02^{+0.13}_{-0.02}$ \\
(1.03,1.33)&$0.45^{+0.07}_{-0.05}$  &$12.61^{+0.34}_{-0.25}$  &$14.19^{+2.44}_{-1.63}$  &$1.05^{+0.03}_{-0.03}$  &$0.99^{+0.02}_{-0.02}$ \\
(1.83,1.88)&$-1.27^{+0.26}_{-0.28}$  &$27.10^{+8.46}_{-5.42}$  &$232^{+185}_{-109}$  & $1.14^{+0.16}_{-0.13}$  &$0.91^{+0.09}_{-0.09}$ \\
(2.38,2.48)&$-0.71^{+0.26}_{-0.31}$  &$18.30^{+3.80}_{-3.21}$  &$43.5^{+33.8}_{-21.5}$  &$0.98^{+0.14}_{-0.11}$  &$0.95^{+0.09}_{-0.08}$ \\
\bottomrule\label{Model fit results for SGR 22110916 fit with High-Gain}
\end{tabular*}
\footnotesize{GRD01,05 represent the constant factors of the respective detectors\\}
\end{table*}
\vspace{-2.0em}
\end{center}

\begin{center} 
\begin{table*}
\vspace{-2.0em}
\setlength{\tabcolsep}{0.009pt}
\caption{Model fit results for SGR 22101215}
\begin{tabular*}{\hsize}{@{}@{\extracolsep{\fill}}ccccccc@{}}
\hline
\midrule
Time range (s) &$\alpha$ & $E_{\rm cut}$ (keV) & norm &  GRD07 & GRD09 & GRD10\\
\midrule
(0.01,0.08) & $0.60^{+0.11}_{-0.11}$  &$11.67^{+0.39}_{-0.46}$  &$11.46^{+3.26}_{-3.39}$  &$0.92^{+0.01}_{-0.05}$  &$0.91^{+0.02}_{-0.03}$  &$0.82^{+0.04}_{-0.04}$ \\
(0.32,0.38) & $1.04^{+0.24}_{-0.13}$  &$9.68^{+0.76}_{-0.37}$  &$3.37^{+2.68}_{-0.89}$  &$1.03^{+0.06}_{-0.05}$  &$0.93^{+0.06}_{-0.05}$  &$0.83^{+0.06}_{-0.04}$ \\
\bottomrule\label{Model fit results for SGR 22101215 fit with High-Gain}
\end{tabular*}
\footnotesize{GRD07,09,10 represent the constant factors of the respective detectors\\}
\end{table*}
\vspace{-2.0em}
\end{center}

\section{EFFECT OF THRESHOLD ON EFFICIENCY}
The trigger threshold is commonly implemented in the electronics to exclude the low energy noise. There are two types of trigger threshold: the hard-threshold where the trigger efficiency is a heaviside step function, and the soft-threshold where the trigger efficiency goes from 0\% to 100\% gradually as the signal amplitude increase. The GECAM-C GRDs use a soft trigger threshold as a compromise between lower count rates and the lower energy range. Through our research, it is found that the effect of threshold on the detection efficiency of the detector can be described by a Gaussian CDF function, and the Gaussian parameters are related to the threshold and the noise level: 

\begin{equation}
\rm
    CDF(\rm E \mid \mu, \sigma) = \frac{1}{2}\Big(1+erf\big(\frac{(E-\mu)}{\sqrt{2}\sigma}\big)\Big)
\end{equation}

\begin{figure}[H]
   \centering
   \includegraphics[width=0.7\textwidth, angle=0]{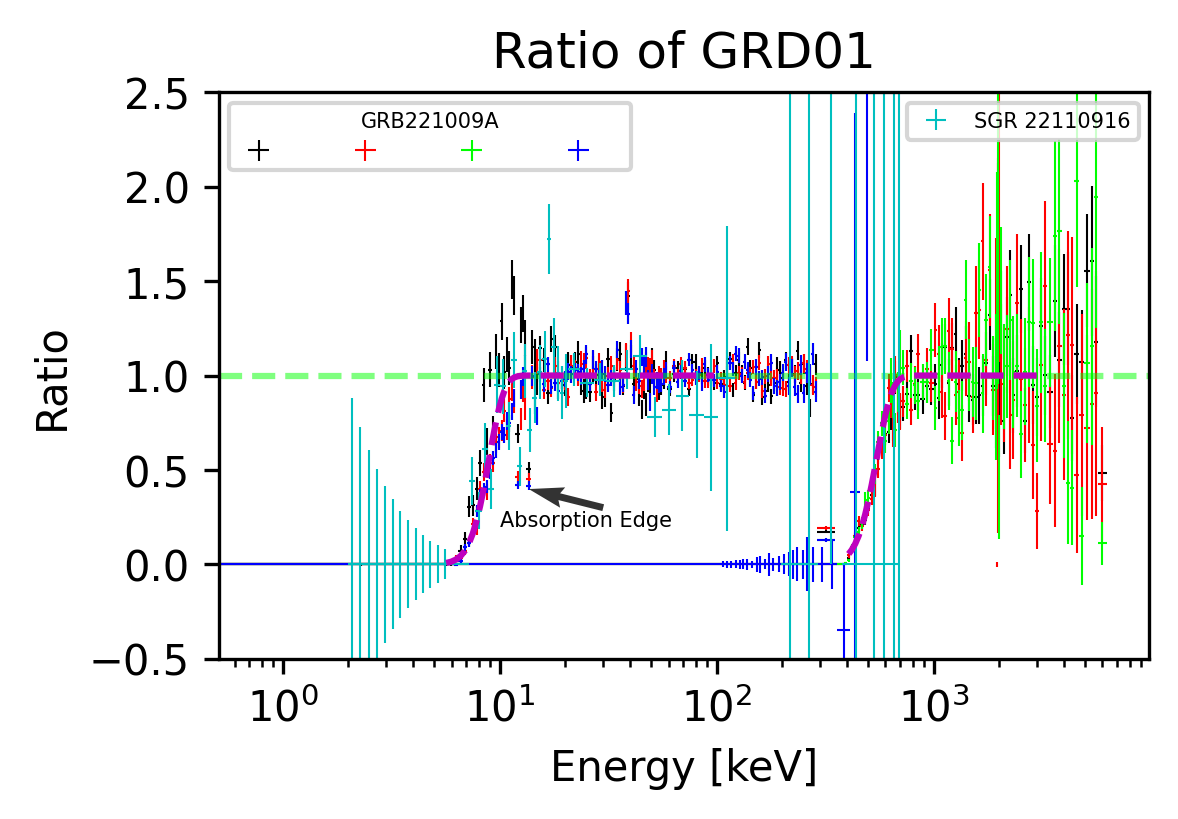}
   \caption{Distribution of the ratio between the data and the model after spectrum fitting for the GECAM-C GRD01 detector; Mainly from two bursts, one GRB and one SGR, the four colored data points under GRB221009A represent the fitting results of the four time segments of this burst;  The left and right CDFs of the graph represent the high and low gains respectively; The magenta dashed line represents the preliminary fit to the data points, we can see that the CDF function fits this curve very well; Moreover, the magnetar and GRB results are consistent.}
   \label{fig:The_Ratio_of_GRD01}
\end{figure}

\begin{figure}[H]
   \centering
   \includegraphics[width=0.7\textwidth, angle=0]{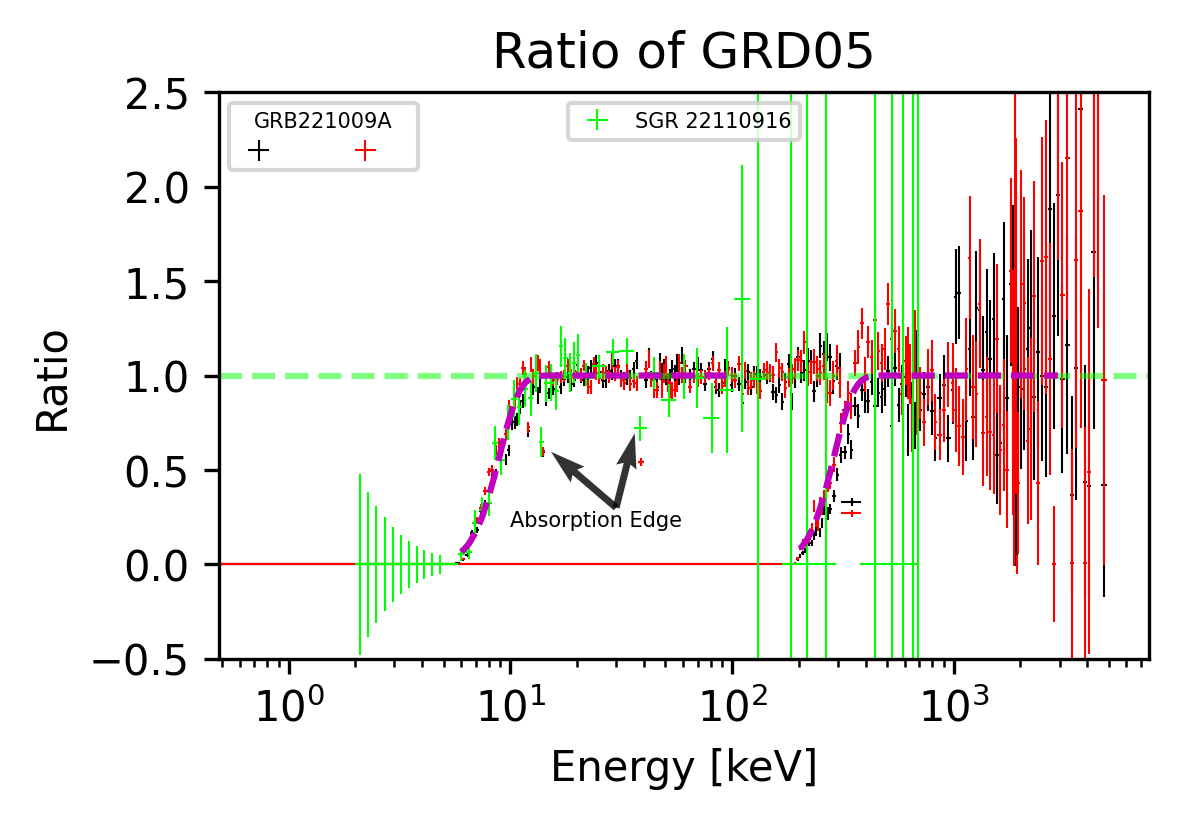}
   \caption{Distribution of the ratio between the data and the model after spectrum fitting for the GECAM-C GRD05 detector; Other is similar to the Fig \ref{fig:The_Ratio_of_GRD01}.}
   \label{fig:The_Ratio_of_GRD05}
\end{figure}

As shown in Fig \ref{fig:The_Ratio_of_GRD01} and Fig \ref{fig:The_Ratio_of_GRD05}, the ratio is defined as the net number of photons deposited at the detector by the source divided by the number of photons incident from the source, and this characterizes the effective area or detection efficiency of the detector to a certain extent. However, due to the effect of the threshold and noise we can see that there is a changing trend at the low energy end of both high and low gains, rather than the 100 percent detection efficiency. In addition, this trend is well described by the Gaussian CDF, which means that we can modify the effect of the threshold on the detection efficiency:

\begin{equation}
\rm
    F_{detector}(E) = RSP_{old} \otimes N_{source}(E) \otimes RSP_{correct}
\end{equation}

where $\rm RSP_{old}$ is the existing corresponding matrix that mainly considers the energy resolution and detection efficiency, $\rm RSP_{correct}$ is the correction component of the threshold on the detection efficiency, and it's the inverse of the CDF.

\begin{figure}[H]
   \centering
   \includegraphics[width=0.62\textwidth, angle=0]{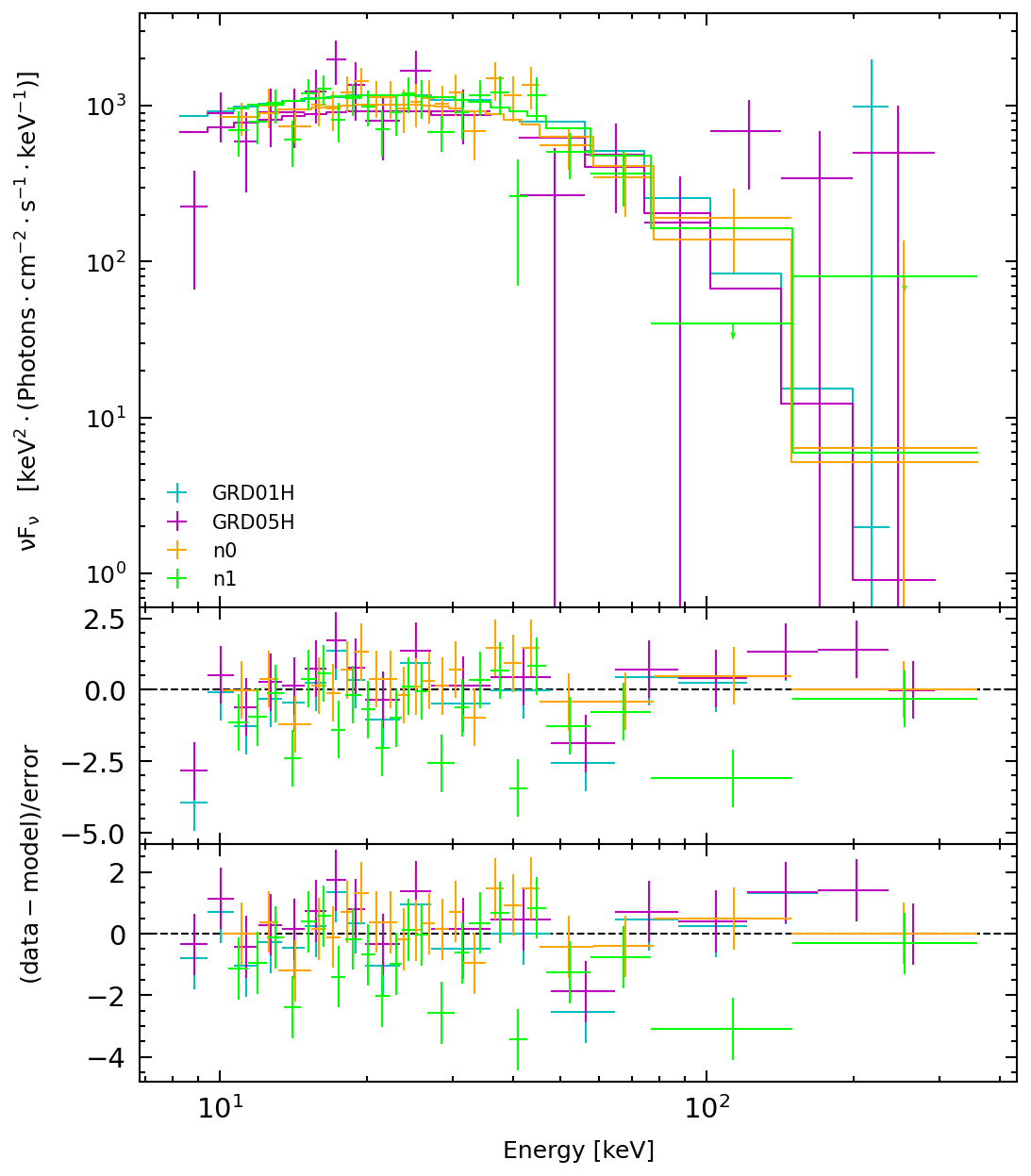}
   \caption{Joint data analysis of GECAM-C and \emph{Fermi}/GBM for SGR22110916; The first subfigure is the energy flux $\rm \nu \mathcal {F}_{\nu}$; The second subgraph is the residuals in terms of sigmas with error bars of size one with the $\rm RSP_{old}$, it is clear that the data points in cyan and magenta deviate significantly from 0 due to the absence of efficiency correction for the low-energy end; The third subgraph is the residuals with the correction, it is clear that the two data points at the low end have been corrected well, because they are closer to zero, which means that the data fit the model better. In addition, the fluctuation at the high-energy end is caused by statistical fluctuations.}
   \label{fig:Spectrum_of_SGR22110916}
\end{figure}

We tried this correction for GRD01 and GRD05 as discussed above, and the effect of the correction can be shown in the Fig \ref{fig:Spectrum_of_SGR22110916}. As can be seen from the residual comparison plot, the effect of correction is very good, this can help us make better use of the GECAM-C scientific data in detectable energy range.

\section{CONCLUSION AND SUMMARY}
In this paper, we used two relatively bright GRBs and some SGRs with joint observations for the cross-calibration work of the GECAM-C GRD detectors, respectively. For GRB220921A, we calibrate both the High-Gain and Low-Gain channel of the GRDs. For the other samples, due to the soft spectrum, we only use the High-Gain of GECAM-C. All calibration results show that the constant factors for the GECAM-C GRDs are close to one, and the spectral model can also be well constrained. These results demonstrate that there is a very good agreement between GECAM-C GRDs and the other well calibrated instruments (i.e. \emph{Fermi}/GBM and \emph{Swift}/BAT). Meanwhile, we also find good agreement between GRD detectors of GECAM-C.

Besides, we try to correct the effect of the threshold of GRD01,05 on the detection efficiency with a Gaussian CDF function. As shown in Fig \ref{fig:Spectrum_of_SGR22110916}, the effect of correction is reasonably good. This can help us make better use of the data at the low energy range (8-10 keV at present) in our analysis.


\begin{acknowledgements}
We thank supports from the staff of the Key Laboratory of Particle Astrophysics, Center for Space Science who offered great help in the development and calibration tests of GECAM-C. 
This research is supported by National Key R\&D Program of China (Grant No. 2021YFA0718500), the Strategic Priority Research Program of Chinese Academy of Sciences (GranNo. XDA15360102), the National Natural Science Foundation of China (Grant No. 12273042).


\end{acknowledgements}

\newpage

-----------------------------------------------------------------------------
\bibliographystyle{unsrt}
\bibliography{sample}

\label{lastpage}
 
\newpage
\begin{appendix}
\section{Thre light curves of GECAM-C}
\setlength{\tabcolsep}{0.009pt}
\subsection{GRB 220921A}
\begin{figure}[H]
   \centering
   \includegraphics[width=0.9\textwidth, angle=0]{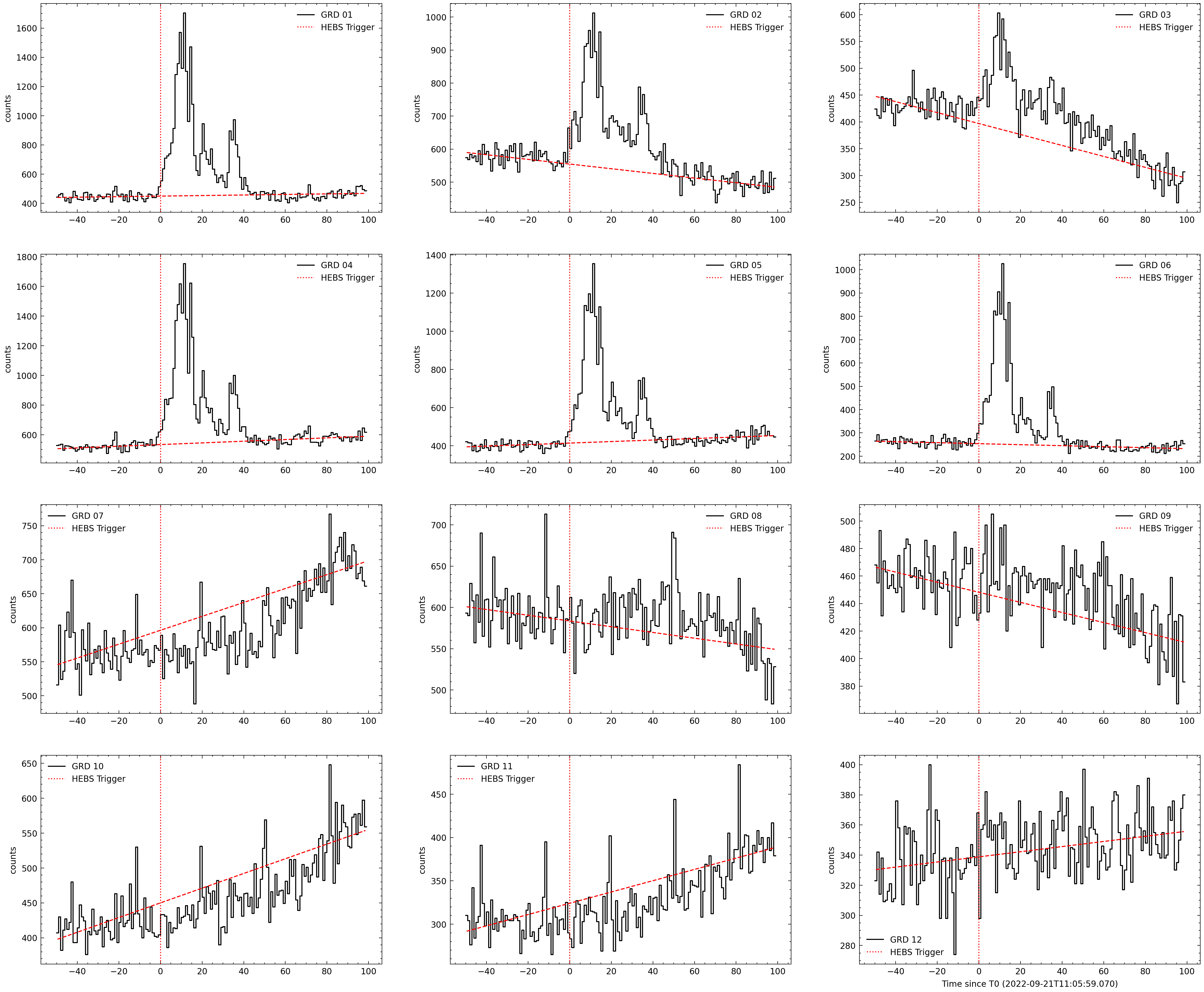}
   \caption{The light curves detected by GECAM-C of GRB 220921A; And the abscissa is the time since the trigger time 2022-09-21T11:05:59.070 (UTC); The twelve subplots represent the light curves of 12 GRDs, respectively.}
   \label{fig:GRB220921A_GECAM-C}
\end{figure}

\begin{figure}[H]
   \centering
   \includegraphics[width=0.6\textwidth, angle=0]{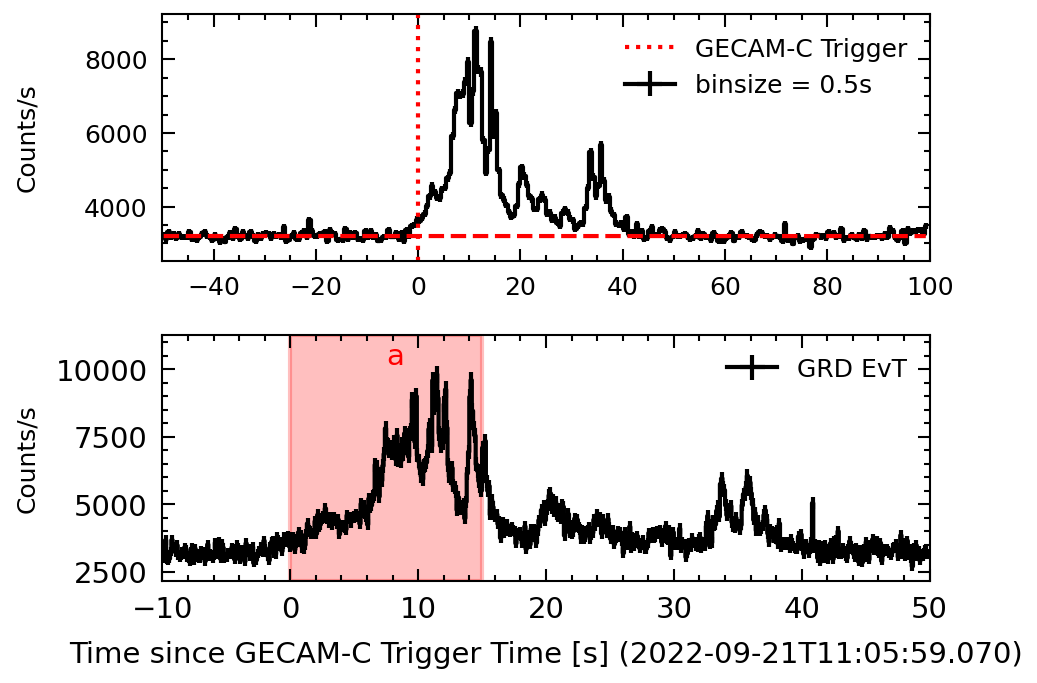}
   \caption{GECAM-C total light curve of GRB220921A and the time interval selection for joint calibration; The first subgraph is the total light curve of the signal detectors (GRD 01,02,03,04,05,06), and the red shaded interval of the second subgraph is the time range (T0+[0,15] s) of the joint energy spectrum fitting.}
   \label{fig:GECAM-C_lightcurve_GRB220921A}
\end{figure}

\subsection{GRB 220930A}
\begin{figure}[H]
   \centering
   \includegraphics[width=0.9\textwidth, angle=0]{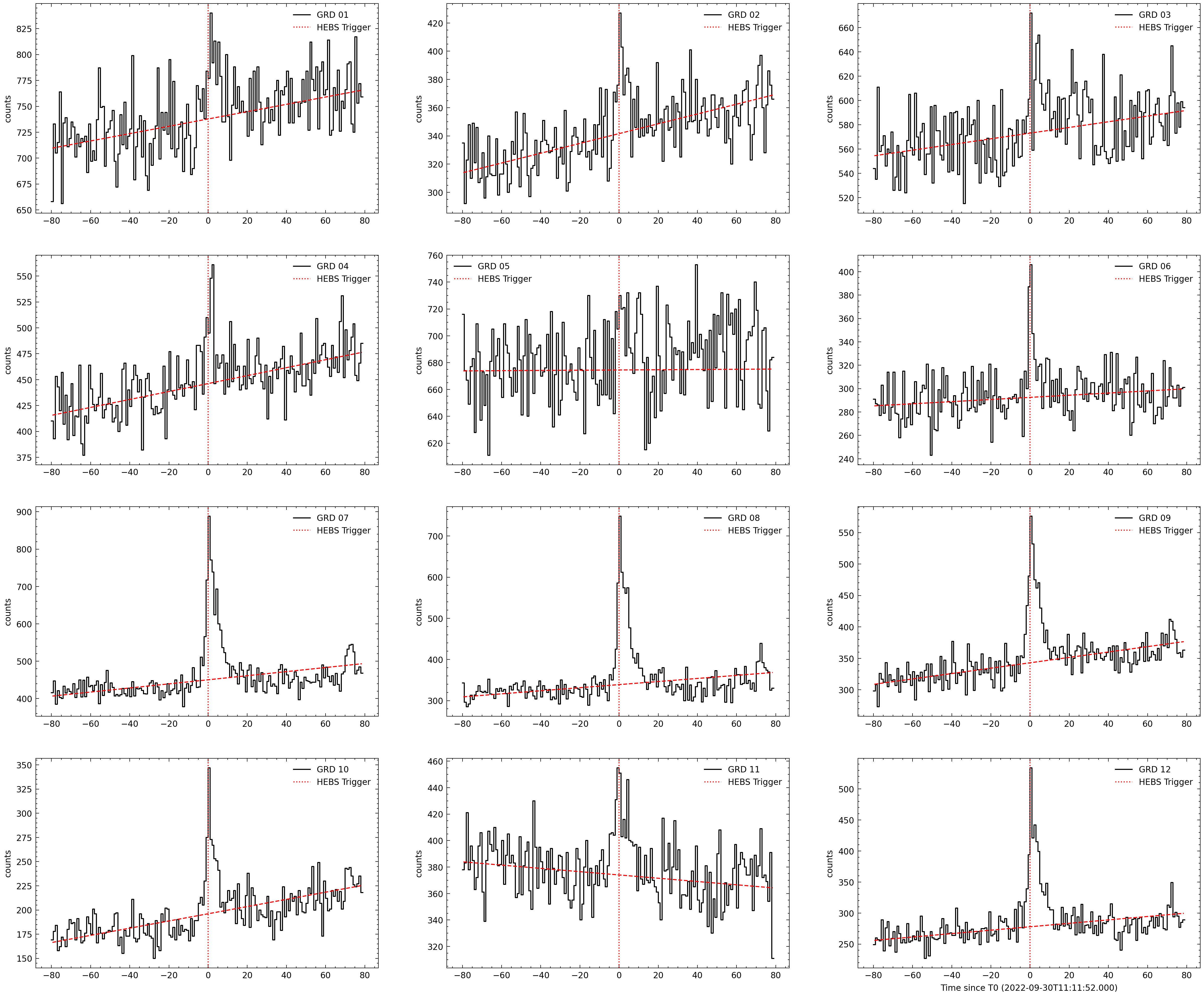}
   \caption{The light curves detected by GECAM-C of GRB 220930A; And the abscissa is the time since the trigger time 2022-09-30T11:11:52.000 (UTC); The twelve subplots represent the light curves of 12 GRDs, respectively.}
   \label{fig:GRB220930A_GECAM-C}
\end{figure}

\begin{figure}[H]
   \centering
   \includegraphics[width=0.6\textwidth, angle=0]{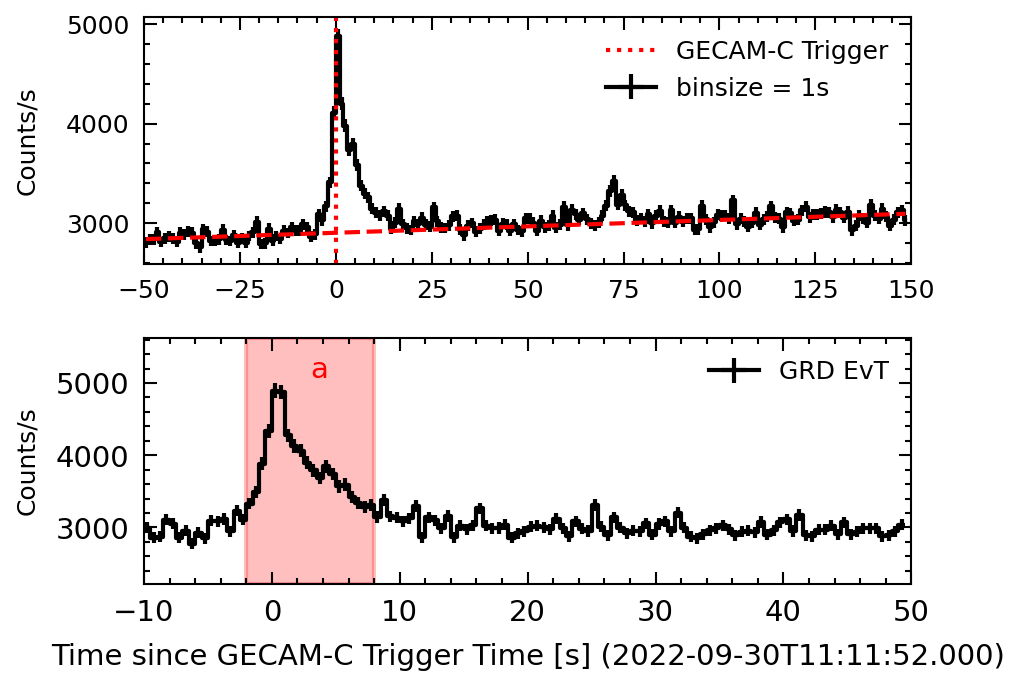}
   \caption{GECAM-C total light curve of GRB220930A and the time interval selection for joint calibration; The first subgraph is the total light curve of the signal detectors (GRD 06,07,08,09,10,11,12), and the red shaded interval of the second subgraph is the time range (T0+[-2,8] s) of the joint energy spectrum fitting.}
   \label{fig:GECAM-C_lightcurve_GRB220930A}
\end{figure}

\subsection{SGR 22110916}
\begin{figure}[H]
\centering
    \begin{minipage}[t]{0.6\linewidth}
    \centering
    \includegraphics[width=0.9\linewidth]{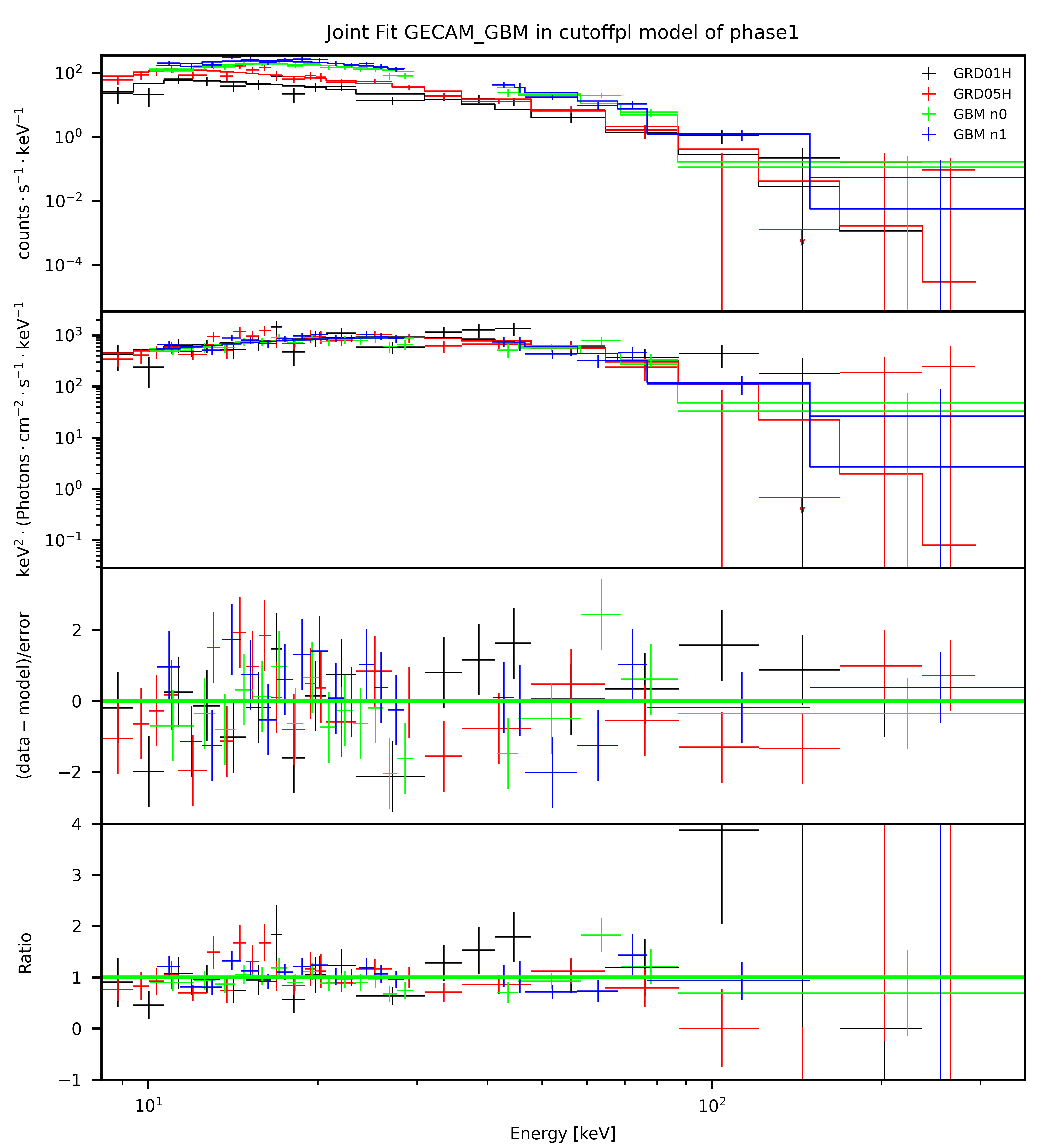}
    \end{minipage}
    \begin{minipage}[t]{0.6\linewidth}
    \centering
    \includegraphics[width=0.9\linewidth]{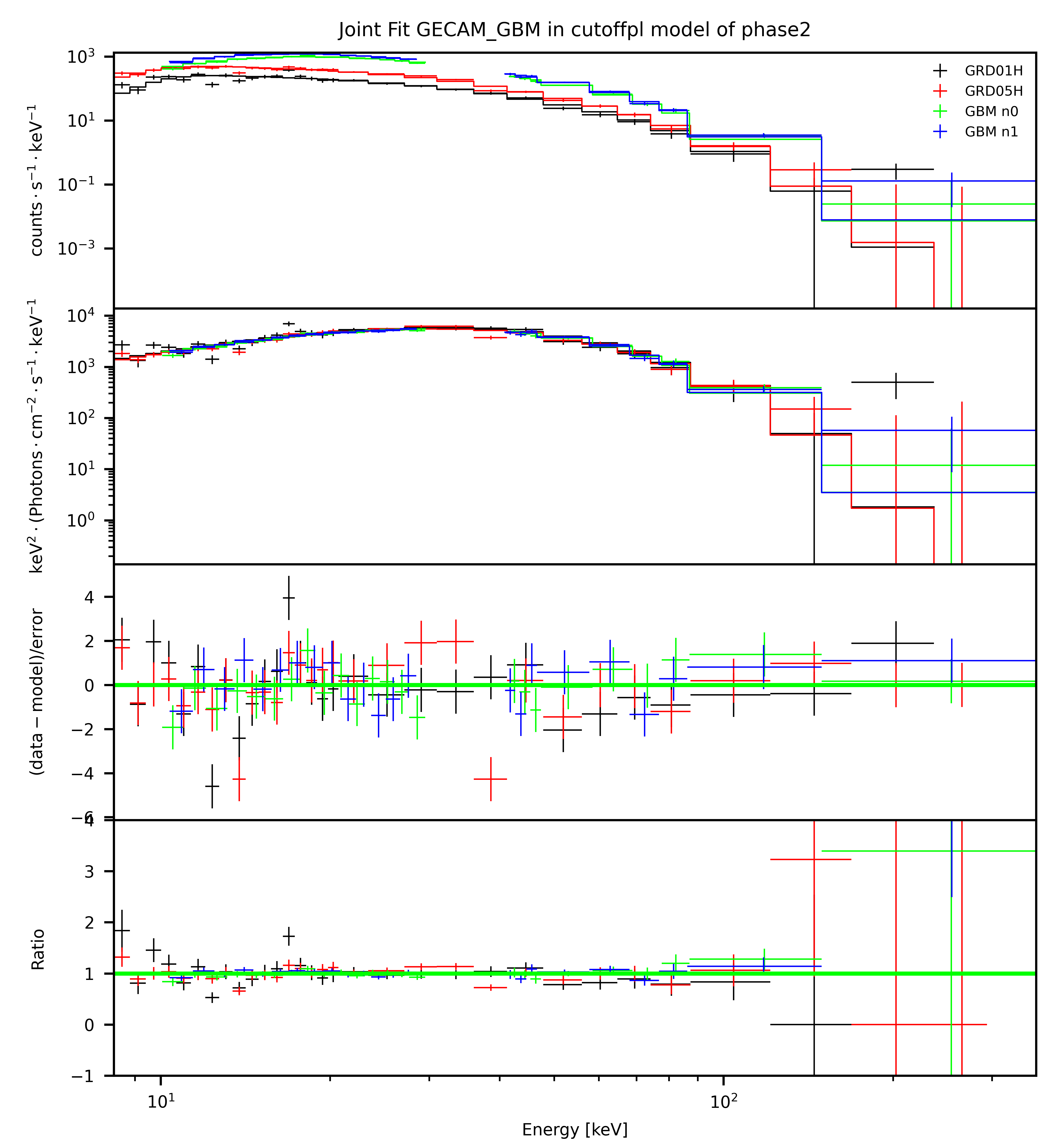}
    \end{minipage}%

    \begin{minipage}[t]{0.6\linewidth}
    \centering
    \includegraphics[width=0.9\linewidth]{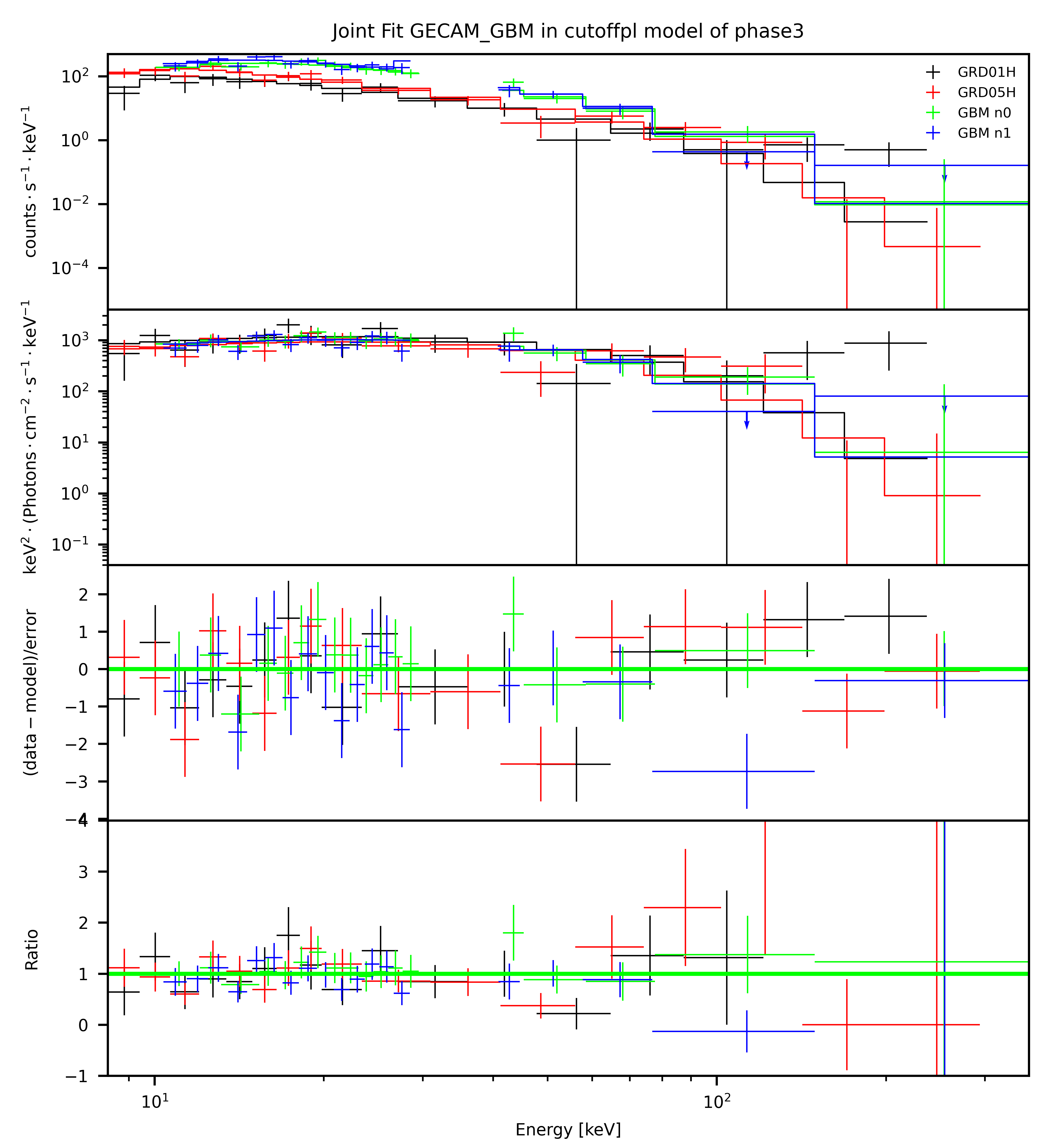}
    \end{minipage}
    \begin{minipage}[t]{0.6\linewidth}
    \centering
    \includegraphics[width=0.9\linewidth]{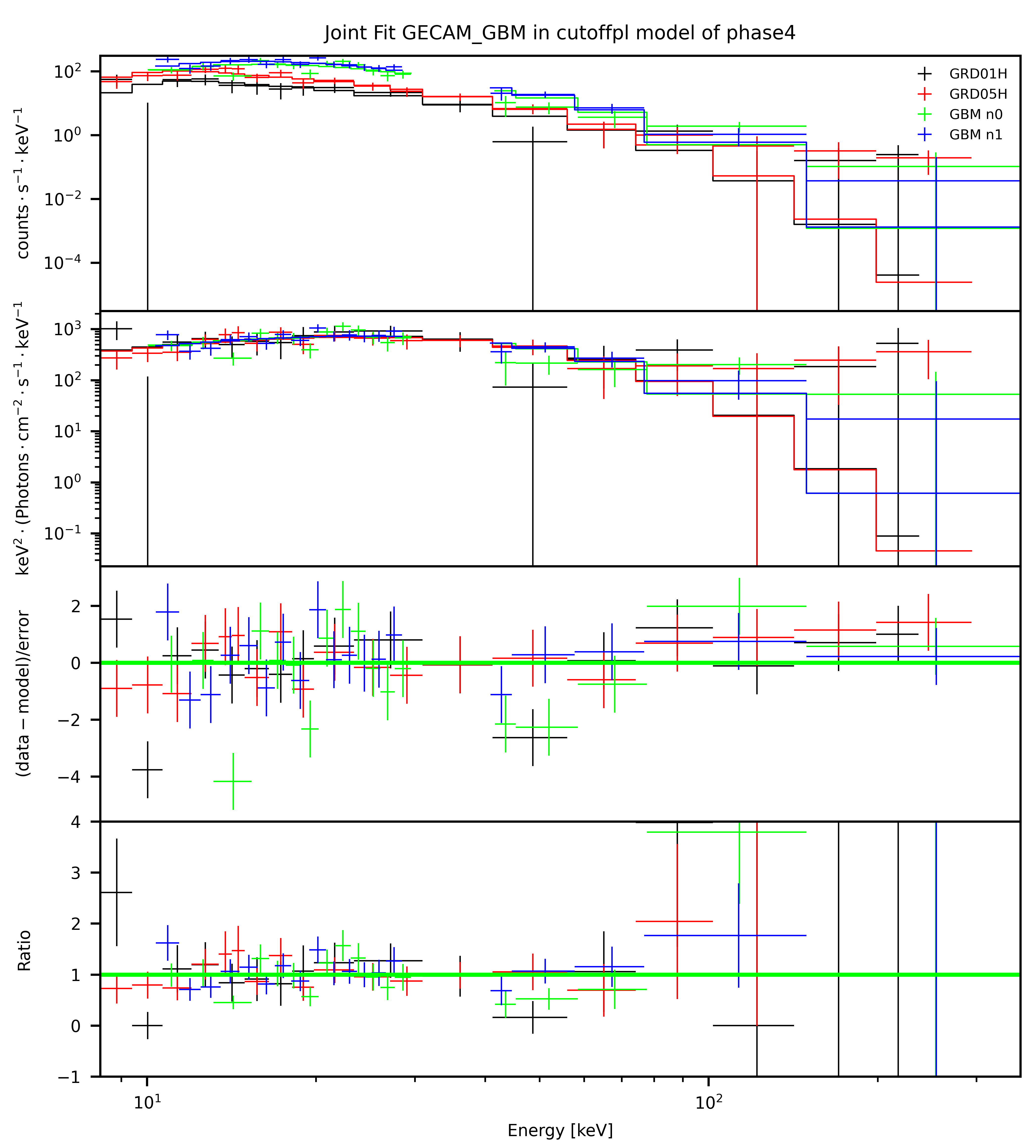}
    \end{minipage}%

\caption{Spectrum fitting results for the four time segments of SGR 22110916}
\label{fig:GECAM-C_spectrum_result_of_SGR22110916}
\end{figure}

\subsection{SGR 22101215}
\begin{figure}[H]
\centering
    \begin{minipage}[t]{0.6\linewidth}
    \centering
    \includegraphics[width=0.9\linewidth]{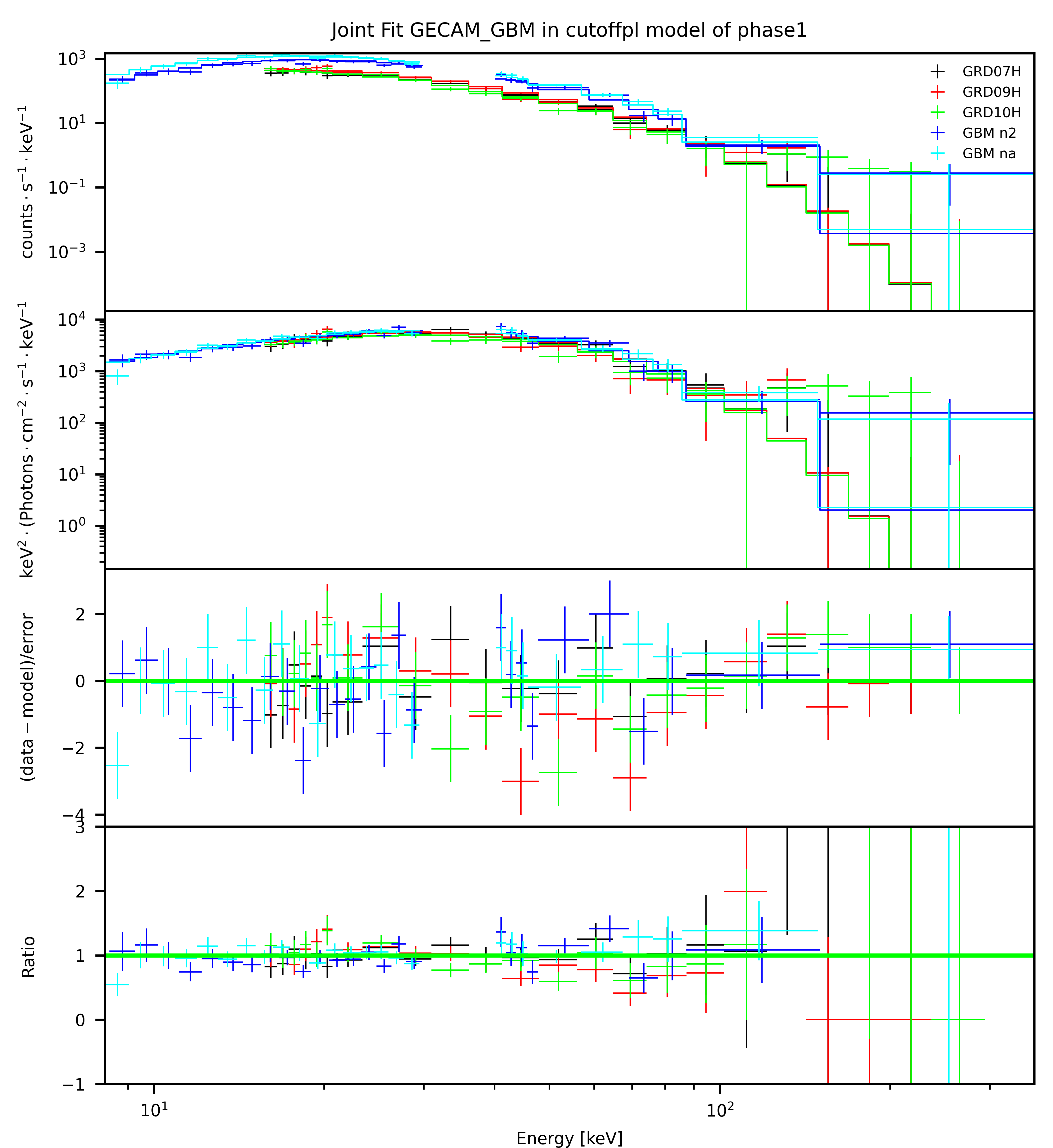}
    \end{minipage}
    \begin{minipage}[t]{0.6\linewidth}
    \centering
    \includegraphics[width=0.9\linewidth]{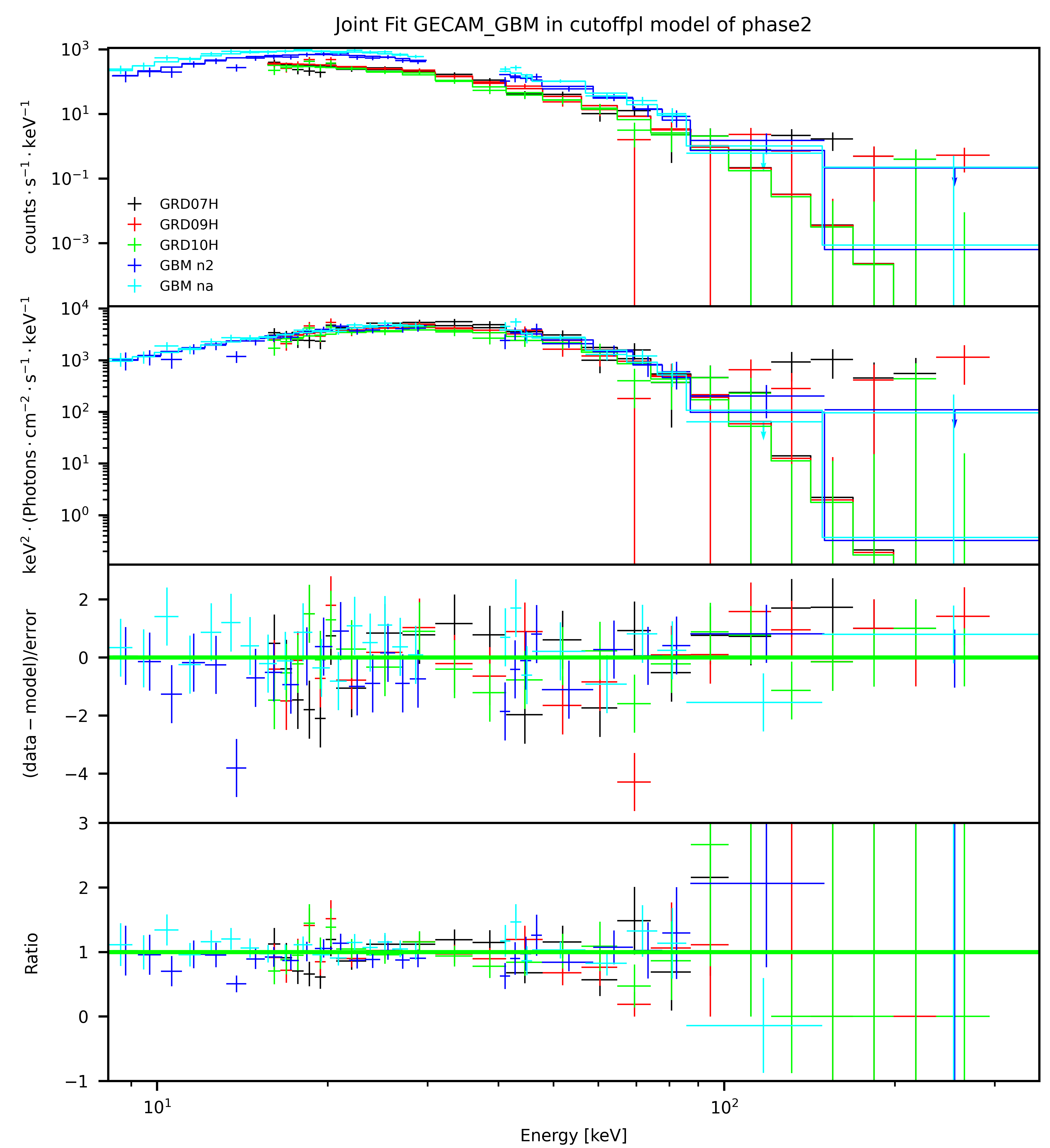}
    \end{minipage}%
\caption{Spectrum fitting results for the two time segments of SGR 22101215}
\label{fig:GECAM-C_spectrum_result_of_SGR22101215}
\end{figure}

\end{appendix}

\end{document}